\documentstyle[aps,prl,preprint]{revtex}
\input{psfig}
\begin{document}
\draft

\title{The onset of magnetic order in fcc-Fe films on Cu(100)}

\author{S S A Razee$^1$, J B Staunton$^1$, L Szunyogh$^2$,
          and B L Gyorffy$^3$}

\address{$^1$Department of Physics, University of Warwick,
               Coventry CV4 7AL, UK}

\address{$^2$Department of Theoretical Physics,
         Budapest University of Technology and Economics, \\
         Budafoki \'ut. 8, H-1521 Budapest, Hungary, and \\
         Center for Computational Materials Science,
         Technical University of Vienna,
         Getreidemarkt 9/158, A-1060 Vienna, Austria}

\address{$^3$H.H. Wills Physics Laboratory, University of
         Bristol, Tyndall Avenue, \\
         Bristol BS8 1TL, UK}

\date{\today}
\maketitle

\begin{abstract}
On the basis of a first-principles electronic structure theory
of finite temperature metallic magnetism
in layered materials, we investigate
the onset of magnetic order in thin (2-8 layers)
fcc-Fe films on Cu(100) substrates. The nature of this ordering is
altered when the systems are capped with copper. Indeed
we find an oscillatory dependence of the Curie temperatures
as a function of Cu-cap thickness, in excellent
agreement with experimental data. The thermally induced spin-fluctuations
are treated within a mean-field disordered local moment (DLM) picture
and give rise to layer-dependent `local exchange splittings' in the
electronic structure even in the paramagnetic phase.
These features determine the magnetic intra- and interlayer
interactions which are strongly influenced by the presence
and extent of the Cu cap.

\end{abstract}




The Disordered Local Moment (DLM) picture \cite {jh+hh+dme}, when
implemented by the KKR-CPA method \cite{blg}, is a first-principles,
that is to say material
specific and yet parameter-free, mean field theory of metallic
magnetism in the limit where good local moments form. It works well
for bulk $Fe$ and $Co$ and their alloys but appears to be inadequate
for $Ni$ \cite{jbs,JBS+BLG}. Under these circumstances it is of fundamental interest
to test its applicability by using it in different, but nevertheless
similar, physical situations such as in a thin films whose thickness, $d$,
can act as a tuneable, new thermodynamic variable. With this in mind
we deployed the theory for $Fe$ on and embedded in $Cu$. In this
letter we report some of our most significant findings.

For complexity in magnetic systems ultrathin $Fe$ films on $Cu(100)$
are unsurpassed \cite{Blugel15}. How the films grow, their structure and morphology
are profoundly intertwined with their magnetic properties and a
satisfactory description of these systems has become a benchmark for
theories of thin film magnetism. After much extensive experimental work
on thermally deposited films it is now clear that
below a critical thickness of 10-12 monolayers (ML) the films take on
the fcc structure of the substrate while thicker
films revert to the bcc structure of bulk iron. In the ultra
thin regime when there are fewer than 4-5 ML a ferromagnetic (FM)
phase is observed whilst the thicker films of 6-11 ML seem to be
antiferromagnetic (AF) with a net moment across the film. Adding
covering layers of $Cu$ to the films has a marked effect upon
their magnetic properties. A single $Cu$ monolayer suppresses the magnetic
ordering temperatures $T_c$ whilst the $T_c$'s of 2ML Cu-capped
films are partly restored. In fact $T_c$ oscillates weakly
as further Cu layers are added \cite{Vollmer}.

The experimental data on the uncapped films has been interpreted \cite{Camley}
in terms of models in which the top two surface iron layers are coupled
together ferromagnetically in an otherwise AF $Fe$ film. This picture
has been prompted by a series of ab-initio $T=0$K electronic structure
total energy calculations \cite{Blugel222324,Sz,Blugel}. Pajda et al.\cite{Pajda}
have calculated effective exchange interactions in a $Cu$-capped $Fe$
monolayer on $Cu(100)$ and obtained an oscillatory variation of $T_c$ with cap
thickness. Here we describe the full DLM picture of both electronic
and magnetic structure
pertinent to the paramagnetic states of $Fe$-films on $Cu(100)$
of up to 8ML thickness at finite temperatures and
investigate both the growth of magnetic correlations and onset of magnetic
order. Moreover we are able to extract intra- and interlayer
`exchange' interactions.

The DLM picture is based on assuming a separation
between fast and slow degrees of freedom in the interacting many-electron
system. For times, $\tau$,
long in comparison with an electronic hopping times,$\hbar/W$
($\approx 10^{-15}$ seconds), where $W$ represents a relevant band-width, but
short when compared with typical spin fluctuation times, the spin
orientations of the electrons leaving an
atomic site are sufficiently correlated with
those arriving that the magnetisation integrated over a unit cell and
averaged over $\tau$ is non-zero. These are the `local moments' which can
change their orientations, described by a set of unit vectors
 $\{ \hat{e}_{i} \}$ labelled by the site index $i$, on a time scale
longer than $\tau$
while their magnitudes fluctuate rapidly on the time scale $\tau$.
This timescale demarcation has been invoked in much recent work
\cite{Kubler,Liech,Schif,Pajda2} on bulk systems but has not been
examined and tested in thin film systems. This is the issue
we address in this letter.

For an ab-initio implementation of this picture, standard Spin Density Functional
theory for studying electrons in spin-polarised metals is adapted to
describe the states of the system for each orientational configuration $\{
\hat{e}_{i} \}$. At the heart of the theory is the generalised electronic
Grand Potential $\Omega \{ \hat{e}_{i}\}$ of the system so constrained.
A long time average can be replaced by an ensemble
average with respect to the Gibbsian measure $P \{ \hat{e}_{i} \} = Z^{-1} \exp
-\beta \Omega \{ \hat{e}_{i}\}$, where the partition function $Z =
\prod_{i} \int d \hat{e}_{i} \exp -\beta \Omega \{ \hat{e}_{i} \}$. The
thermodynamic free energy, which accounts for the entropy associated with the
orientational fluctuations as well as creation of electron-hole pairs, is
given by $F= -k_{B} T \log Z$. Evidently $\Omega \{\hat{e}_{i}\}$ plays the role
of a classical `spin' (local moment) Hamiltonian, albeit a highly complicated one.
By choosing a suitable reference single-site `spin'
Hamiltonian  $\Omega_{0}\{ \hat{e}_{i}\} = \sum_{i} \omega_{i}(\hat{e}_{i})$
and using the Feynman-Peierls' inequality \cite{Feynman} a mean field theory
is constructed \cite{blg,JBS+BLG}. This is a `first principles' formulation of the
Disordered Local Moment (DLM) picture which
can be implemented by an adaptation of
SCF-KKR-CPA method ideally suited for calculating the partial averages
$\omega_{i}(\hat{e}_{i})$. The approach can be
further improved via the construct of a generalised Onsager cavity field
\cite{JBS+BLG}.

When the above procedure was used to study bulk magnetic metals, it was found
that the local electronic structure can possess a `local exchange'
 splitting even in the paramagnetic
state \cite{blg}. This means that an electron spin-polarised
parallel to a local moment will have a different density of states to that
polarised anti-parallel. When all orientations of the moments are averaged over
in the paramagnetic state the electronic structure is inevitably unpolarised
but consequences from the presence of local moments can still be identified.
  (Moreover the magnitude of this splitting is expected to vary sharply as a
  function of wave-vector and energy. At some points, if the `bands' are
  flat in a region of wave-vector and energy space, the local exchange
splitting will be roughly of the same size as the rigid exchange splitting
of the majority and minority-spin
bands of the ferromagnetic state whereas at other points, where the
  `bands' have greater dispersion, the splitting will vanish entirely.)
  This local exchange-splitting is the cause
  of the establishment of a local moment.
Photoemission \cite{Kisker} and inverse photoemission (IPES) experiments
\cite{Kirschner} have found these qualitative features in bcc $Fe$. In our
generalisation of the DLM theory to layered systems we find
that the layer dependence of similar features in the electronic structure
of paramagnetic $Fe$ grown on a $Cu(100)$ substrate drives the onset of
magnetic order and the form of the magnetic interactions between $Fe$ layers.

To probe the content of the theory consider the response of the paramagnetic DLM state
to the application of a small, external spin-only magnetic field,
$ \{ {\bf h_i} \} $, varying from site to site, layer to layer
in both orientation and magnitude. The induced magnetisation is
predominantly caused by the local moments changing their orientations to
align with the field causing a change in the single-site
probabilities, $ \delta P_i ( {\bf e}_i ) $.  In short the site by site
paramagnetic spin susceptibility \cite{jbs} is the solution of the matrix
equation,
\begin{equation}
\chi^{ij} = \frac{\beta}{3} \mu_i^2 \delta_{ij} +
            \frac{\beta}{3} \sum_k S^{(2)}_{ik} \chi^{kj}
  \label{eq:chiij}
\end{equation}
where $S^{(2)}_{ik}$ is the corresponding direct correlation function and
$ \mu_i $ is the magnitude of the local magnetic moment on the
$i$-th site. Expressions for $S^{(2)}_{ij}$ involving the electronic
structure of the paramagnetic state and techniques for calculating them,
using the KKR-CPA, for bulk systems have been given elsewhere \cite{jbs}.
For layered systems with two-dimensional translational symmetry the
magnitudes of the `local moments', $ \{ \mu_i \} $, assume one value
$\mu_{P}$  per site in a given layer labelled $P$ but vary from layer
to layer. Taking a 2D lattice Fourier transform over sites within each
layer equation (\ref{eq:chiij}) can be rewritten as,
\begin{eqnarray}
\chi^{PQ} ({\bf q}_\parallel) & = & \sum_{ij} \chi^{PiQj} \;
      \exp [-i {\bf q}_\parallel \cdot ( {\bf R}_i - {\bf R}_j ) ]
    \nonumber \\
            &  = & \frac{\beta}{3} \mu_P^2 \delta_{PQ} +
            \frac{\beta}{3} \sum_{S}
    S^{(2)}_{P S} ({\bf q}_\parallel) \; \chi^{S Q} ({\bf q}_\parallel)
  \label{eq:chipq}
\end{eqnarray}
where, $ {\bf q}_\parallel $ is a wave-vector in the 2D layer Brilliouin Zone.
Once the $S^{(2)}_{P S} ({\bf q}_\parallel)$'s have been obtained from the
SCF-KKR-CPA calculations and loaded into a $n \times n$ matrix,
$S^{(2)} ({\bf q}_\parallel)$ (where $n$ is the number of layers in the film),
 this set of equations (\ref{eq:chipq}) is solved
by a simple matrix inversion i.e.$
\chi^{PQ} ({\bf q}_\parallel) = \left[ 3 k_B T I -
   S^{(2)} ({\bf q}_\parallel) \right]^{-1}_{PQ} \mu^2_Q$ ($I$ is a unit
matrix). For films in which the intralayer exchange coupling is ferromagnetic,
as for the $Fe$ films on $Cu$,
the Curie temperature $T_c$ is specified by the condition
$  \| 3 k_B T_C I - S^{(2)} ({\bf q}_\parallel=0) \| = 0$ or in terms of the
largest eigenvalue of $ S^{(2)} ({\bf q}_\parallel=0)$. Full technical
details on how $ S^{(2)} ({\bf q}_\parallel)$ is calculated for layered
systems will be provided elsewhere.

We use the spin-polarised screened KKR method \cite{ls1} for layered systems
adapted for the DLM picture. The lattice
constants of the layers are taken to be the same as that
of the substrate (6.83 a.u.) so that the effects of lattice strain on the
electronic structure are neglected. For each $ n $, the electronic
structure of the DLM state is calculated self-consistently
using 78 $ k_\parallel $ points in the irreducible part of
the surface Brillouin zone.
A buffer of three layers of the
substrate as well as a buffer of at least three layers of vacuum
is calculated self-consistently along with the
potentials on each layer. The self-consistent layer-resolved
potentials are then used to calculate the magnetic
moments and the effective 'Weiss' field on each layer
as described above. By introducing
a small change in the average magnetisation on one
particular layer and calculating the effective 'Weiss'
field on each layer we determine $ S^{(2)}_{PQ} ({\bf q}_\parallel=0) $,
and hence the static paramagnetic spin susceptibility,
$ \chi^{PQ} ({\bf q}_\parallel=0) $ together with the Curie
temperatures $ T_c$'s for various thicknesses of both the $Fe$ film
and the $Cu$ cap.

Table \ref{table1} shows our results for the layer-dependent
`local' magnetic moments in uncapped Fe$_{n}$/Cu(100) films together
with the dependence of the Curie temperatures on Fe-film thickness.
The moments on the topmost layer are always the largest at around
2.5$\mu_B$ followed by the moments on the Fe layer closest to the
substrate next at roughly 2.2$\mu_B$. In the interior of the film
the moments are reduced to 1.7$\mu_B$ close to what was found for
fcc-Fe for this lattice spacing \cite{Pinski}. The magnetic ordering
temperatures reduce monotonically to a constant value of 485 K for
6 MLs onwards. Table \ref{table2} shows the effective `exchange'
interactions, that is to say the direct correlation function,$S^{(2)}_{PQ}$
which lead to these transition
temperatures. Examples of a 7ML and a 3ML Fe layer systems are shown
to bring out the dominant features. The intralayer couplings are
ferromagnetic throughout. The layer at the top of film and the
layer nearest the Cu substrate have the largest values owing to their
reduced effective coordination. It is the interlayer couplings that
show an important and interesting trend. The top two layers are strongly
ferromagnetically (FM) coupled (positive values) whereas the coupling
to nearest neighbor layers within the films is anti-ferromagnetic (AF).
There is also significant coupling to subsequent layers which alternates
between FM and AF.

\begin{figure}
\center{\psfig{figure=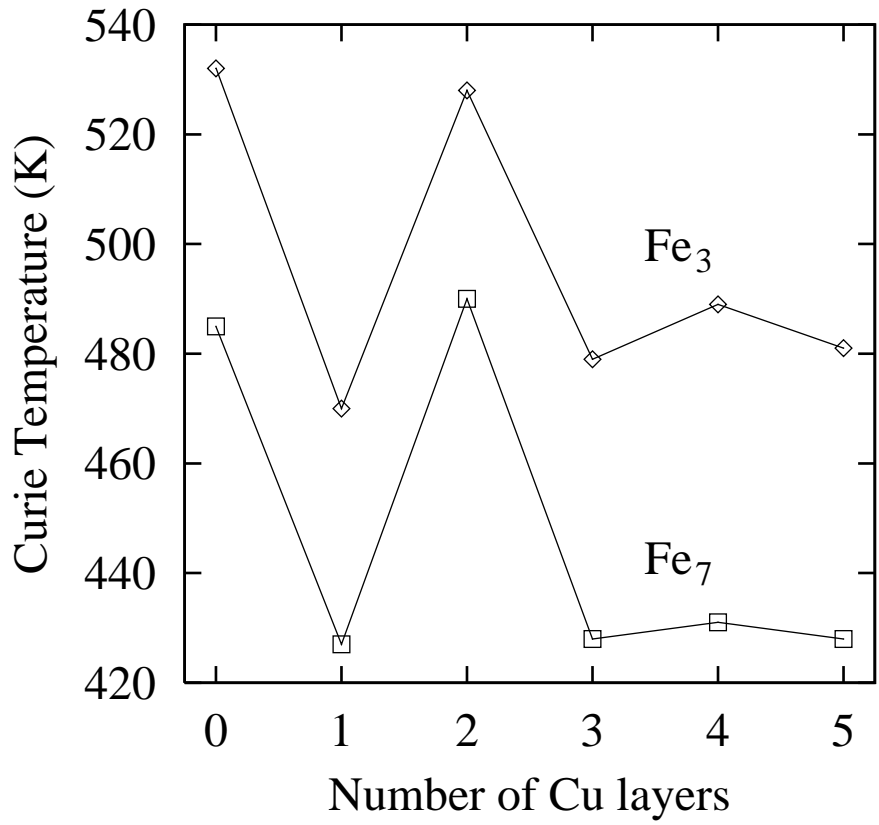}}
\caption{The magnetic ordering temperatures for 3 and 7 layers of
$Fe$ on $Cu(100)$ as a function of number of copper capping layers.}
\label{fig1}
\end{figure}

\begin{figure}
\center{\psfig{figure=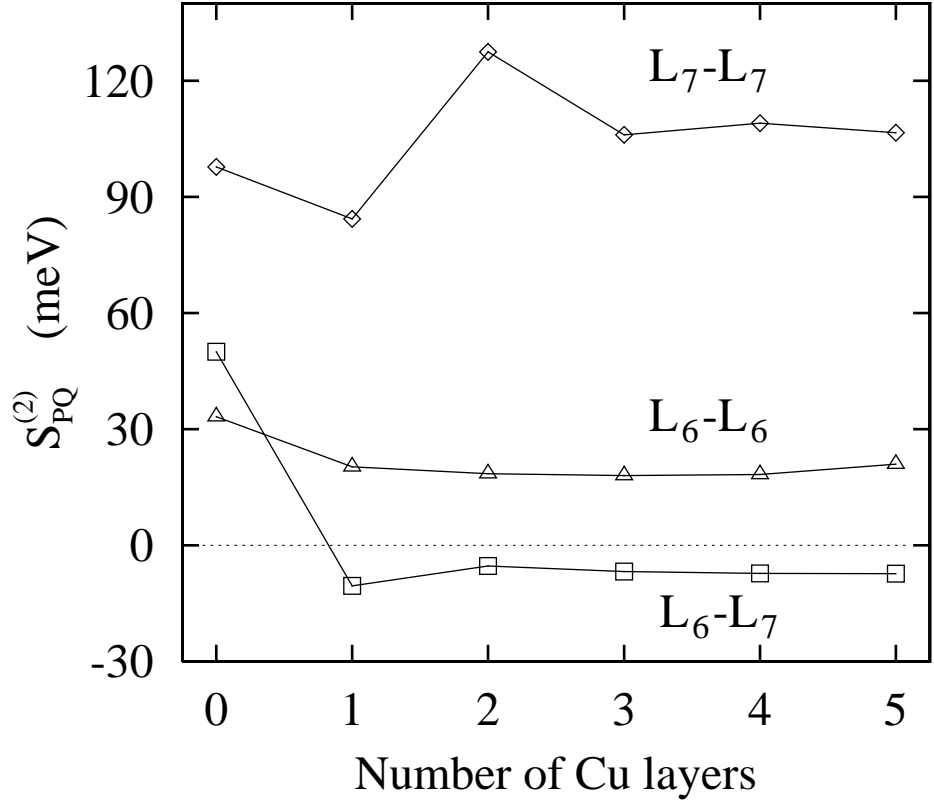}}
\caption{The magnetic interactions in and between the $Fe$ layers
nearest the surface or copper cap in a system of 7 $Fe$ layers on
$Cu(100)$. The values are given in meV and `$L_7$-$L_7$' denotes the
interactions within the top $Fe$ layer (nearest the cap), `$L_6$-$L_6$'
for the second-most top layer and `$L_6$-$L_7$' the coupling between these
two iron layers.}
\label{fig2}
\end{figure}

Adding a copper cap to these films alters these results profoundly.
Figure \ref{fig1} highlights this for 3 and 7 monolayer Fe/Cu(100) films
and shows the magnetic ordering temperature $T_{c}$ to have an
oscillatory dependence on cap thickness. A single Cu monolayer
suppresses $T_{c}$ by some 50K with the deficit being restored
by a second layer. A third layer reduces $T_{c}$ once again and
further layers cause minor oscillations about this lower value.
This behavior is in excellent agreement with experimental data
on the same systems by Vollmer et al.\cite{Vollmer} (see their
figure 6). In figure \ref{fig2} the dominant changes to intra-
and interlayer magnetic interactions are shown. These occur in
the top two Fe layers nearest the cap. A single Cu ML cap switches
the coupling between these layers from FM to AF whilst further
Cu layers strengthen the coupling within the topmost Fe layer.

\begin{figure}
\center{\psfig{figure=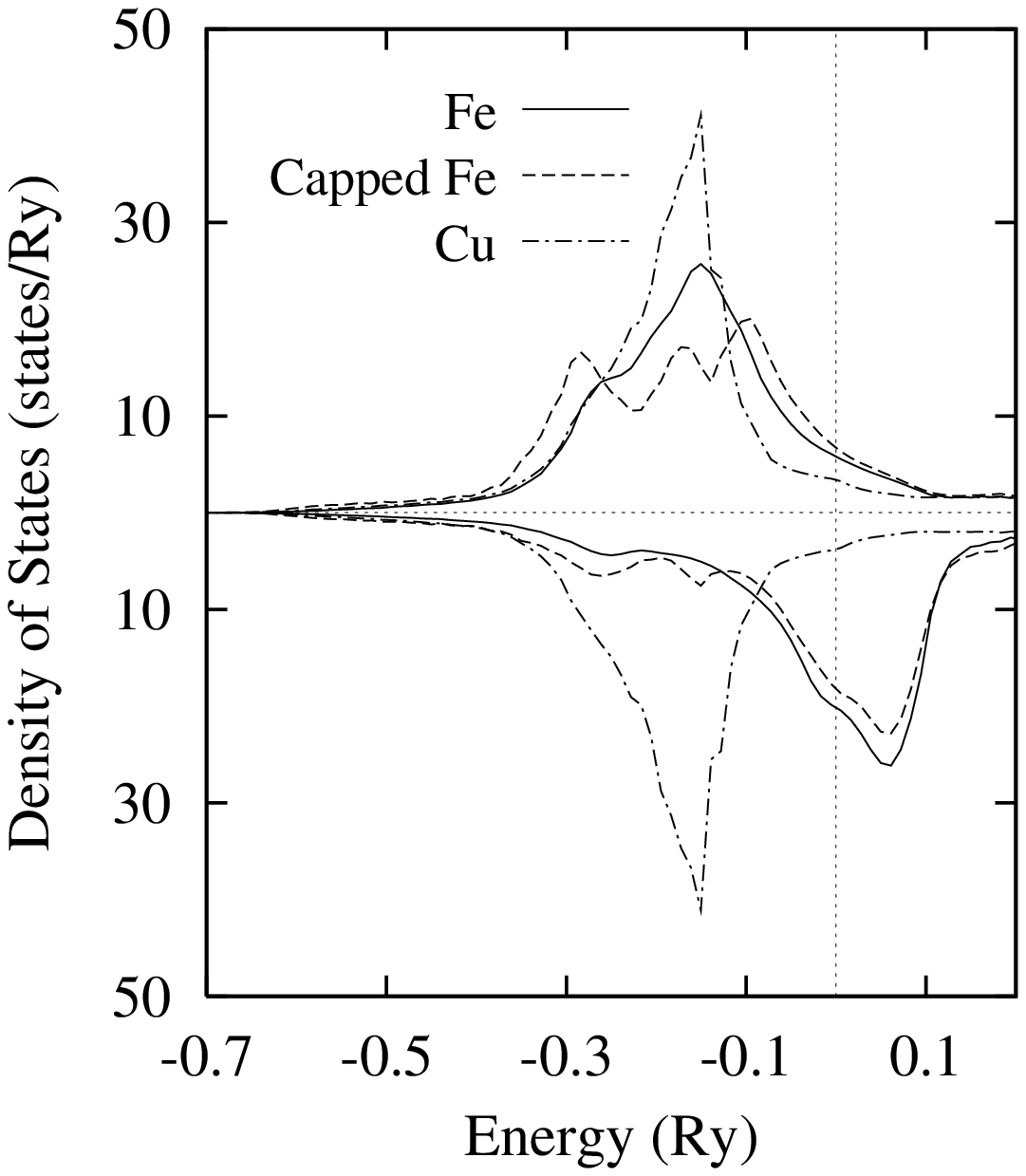}}
\caption{The density of states (DOS) in the topmost Fe layer of
a Fe$_7$Cu(100) system. The full line shows the uncapped
system whilst the dashed line shows the DOS when a single Cu
monolayer caps the system. The dotted line shows the DOS for
this Cu layer. The upper half of the figure shows the DOS for
an electron spin-polarised parallel to the local moment on a site
whereas in the lower half the DOS of an electron polarised
anti-parallel is shown.}
\label{fig3}
\end{figure}

Calculations for bulk fcc-Fe have shown how the tendency to magnetic
order changes as the lattice spacing is varied \cite{Moruzzi,Pinski}.
On an expanded lattice, a ferromagnetically ordered state is stable at
low temperatures and the coupling between `local moments' in the
paramagnetic DLM state mirrors this aspect. As the lattice is contracted
the magnetic ordering tendencies become anti-ferromagnetic.
Capping fcc Fe films with Cu causes a similar effect. Figure \ref{fig3}
shows that the electronic density of states (DOS) of an uncapped film
is `exchange split' over a large energy range. These states
are nearly fully occupied and FM coupling between neighboring sites results.
The reduced coordination of sites within this layer produce an
effect akin to that of increasing the lattice spacing of bulk fcc
Fe. There is hybridisation of these Fe states with quantum well
states in the Cu cap formed between the vacuum and the Fe film.
This causes the exchange splitting to collapse in the region of
this hybridisation just as in fcc Fe for smaller lattice spacings.
The remaining exchange-split states are roughly half-filled which
leads an AF coupling along the (100) direction as in bulk fcc Fe.
Figure \ref{fig3} shows these effects in the DOS in the top layer
of a Fe$_7$/Cu(100) system both uncapped and capped by a single
Cu ML. The narrow band of states in the Cu layer are also shown.

In conclusion we note that the dramatic variations of interaction between
local moments at different distances from the various surfaces and
interfaces are quintessentially itinerant effects. As the much
studied variation of the local moments near such planar defects
show, they are due to the motion of the electrons not limited
to an atom but confined by the geometry of the sample. In this
respect they are manifestations of the same physics as is at work
in the oscillatory magnetic coupling in metallic multilayers \cite{ex-coupl}.
Consequently, the detailed quantitative account of the complex
experimental data by our calculations can be taken as significant
new evidence that the first-principles (KKR-CPA) implementation of
the DLM picture correctly captures the essential physics of magnetic
order, due to mobile electrons, in the limit where local moments
form.

{\it Acknowledgments}:
This work is a result of a collaboration under the
TMR-Network (contract no. ERBFMRXCT 96-0089) on ``{\it Ab initio}
calculations of magnetic properties of surfaces, interfaces,
and multilayers'' and also the RTN 'Computational
Magnetoelectronics' network
(contract number  HPRN-CT-2000-00143) and is supported by the
Engineering and Physical Sciences Research Council (UK) and
the Hungarian National Science Foundation (contracts OTKA
T030240 and T029813).


\begin{table}
\caption{'Local' magnetic moments in $ \mu_B $ on different layers and
the Curie temperatures of uncapped
Fe$_n$/Cu(100) systems. The Fe Layer $ L_1 $ is adjacent to the
substrate and layer $ L_n $ is the top layer.}

\begin{tabular}{|c|cccccccc|c|}
  & \multicolumn{8}{c|}{'Local' moments ($ \mu_B $)} &
$T_{c}$(K)  \\ \tableline
n & $L_1$ & $L_2$ & $L_3$ & $L_4$ & $L_5$ & $L_6$ & $L_7$
& $L_8$ &  \\ \tableline
2 & 2.24 & 2.48 &      &        &        &        &        &
                                & 681  \\
3 & 2.21 & 1.70 & 2.53 &        &        &        &        &
                                & 532  \\
4 & 2.21 & 1.67 & 1.74 & 2.53   &        &        &        &
                               & 495  \\
5 & 2.21 & 1.67 & 1.71 & 1.74   & 2.53   &        &        &
                               & 492   \\
6 & 2.21 & 1.67 & 1.71 & 1.71   & 1.74   & 2.53   &        &
                               & 486  \\
7 & 2.21 & 1.67 & 1.71 & 1.72   & 1.71   & 1.74   & 2.53   &
                                & 485  \\
8 & 2.21 & 1.67 & 1.71 & 1.72   & 1.71   & 1.71   & 1.74   &
                        2.53   & 485  \\ \tableline
\end{tabular}
\label{table1}
\end{table}
\begin{table}
  \caption{Intra- and inter-layer effective `exchange' interactions in meV
 between layers in an uncapped  Fe$_7$/Cu(100) system with the values
for Fe$_3$/Cu(100) shown for comparison in brackets.
  The Fe Layer $ L_1 $ is adjacent to the
  substrate and layer $ L_n $ is the top layer.}
  \begin{tabular}{|c|c|cccc|cc|}
n & $L_1$ & $L_2$ & $L_3$ & $L_4$ & $L_5$ & $L_6$ & $L_7$ \\ \tableline
1 & 108.5 (108.4) & -7.5 &  8.5 & -2.1 &  -0.8 & -0.7 (-10.9) &  -1.0 (25.1) \\ \tableline
2 & -7.5 & 19.6 & -7.5 &  3.2 &  -2.2 & -0.6 & -0.7 \\
3 & 8.5 &  -7.5 & 17.5 & -7.2 &  2.1 & -1.7 & -0.7 \\
4 & -2.1 &  3.2 &  -7.2 &  18.8 &  -8.1 &  2.1 & -2.1 \\
5 & -0.8 & -2.2 &  2.1 & -8.1 & 19.9 & -10.0 &  9.8 \\  \tableline
6 & -0.7(-10.9) & -0.6 & -1.7 &  2.1 & -10.0 & 33.2(29.9) & 50.0 (53.8)\\
7 & -1.0(25.1) & -0.7 & -0.7 & -2.1 &  9.8 & 50.0 (53.8) & 97.7 (96.7) \\
\end{tabular}
\label{table2}
\end{table}

\newpage

\begin{references}

\bibitem{jh+hh+dme} J.Hubbard, Phys. Rev.B {\bf 20}, 4584, (1979);
H.Hasegawa, J. Phys. Soc. Japan {\bf 46} 1504, (1979);
D.M.Edwards, J. Phys.F {\bf 12}  1789, (1982).
\bibitem{blg} B.L.Gyorffy et al. J. Phys. F {\bf 15} 1337,(1985);
J.B.Staunton et al., J. Phys.F {\bf 15}, 1387, (1985).
\bibitem{jbs} J.B.Staunton et al., J. Phys.F {\bf 16} 1761, (1986)
\bibitem{JBS+BLG} J.B.Staunton and B.L.Gyorffy, 1992,
 Phys.Rev.Lett. {\bf 69}, 371.
\bibitem{Blugel15} M.Straub et al., Phys.Rev.Lett. {\bf 77},
743, (1996).
\bibitem{Vollmer} R.Vollmer et al., Phys.Rev.B {\bf 61},
 1303 (2000).
\bibitem{Camley} R.E.Camley and Dongqi Li, Phys.Rev.Lett. {\bf 84},
4709, (2000).
\bibitem{Blugel222324} C.L.Fu and A.J.Freeman, Phys.Rev.B
{\bf 35}, 925, (1987); G.W.Fernando and B.R.Cooper, Phys.Rev.B
{\bf 38}, 3016, (1988); T.Kraft et al., Phys.Rev.B {\bf 49},
11511, (1994).
\bibitem{Sz} R. Lorenz and J. Hafner, Phys. Rev. B {\bf 54}, 15937 (1996);
L. Szunyogh et al., Phys. Rev. B {\bf 55}, 14392 (1997)
\bibitem{Blugel} T.Asada and S.Blugel, Phys.Rev.Lett. {\bf 79},
507, (1997).
\bibitem{Pajda} M.Pajda et al., Phys.Rev.Lett. {\bf 85}, 5424, (2000).
\bibitem{Kubler} M.Uhl and J.Kubler, Phys.Rev.Lett.
 {\bf 77}, 334, (1996).
\bibitem{Liech} A.I.Liechtenstein et al., J.Mag.Magn.Mat.
{\bf 67}, 65, (1987).
\bibitem{Schif} M. van Schilfgaarde and V.P.Antropov,
J.Appl.Phys. {\bf 85}, 4827, (1999).
\bibitem{Pajda2} M.Pajda et al., cond-mat/0007441.
\bibitem{Feynman} R.P.Feynman, Phys.Rev. {\bf 97}, 660, (1955).
\bibitem{Kisker} E.Kisker et al., Phys.Rev.Lett. {\bf 52}, 2285, (1984).
\bibitem{Kirschner} J.Kirschner et al., Phys.Rev.Lett. {\bf 53},
612, (1984).
\bibitem{ls1} L.Szunyogh et al., Phys.Rev.B {\bf 51}, 9552, (1995);
R.Zeller et al., Phys.Rev.B {\bf 52}, 8807, (1995);
L.Szunyogh et al., Phys.Rev.B {\bf 49} 2721 (1994).
\bibitem{Pinski} F.J.Pinski et al., Phys.Rev.Lett. {\bf 56},
{\bf 56}, 2096, (1986).
\bibitem{Moruzzi} C.S.Wang et al., Phys.Rev.Lett. {\bf 54},
1852, (1985); V.L.Moruzzi and P.M.Marcus, `Energy band
theory of metallic magnetism in the elements',{\it Handbook of
Magnetic Materials},{\bf 7}, eds: K.H.J.Buschow (Amsterdam:
North Holland), 97, (1993).
\bibitem{ex-coupl} P.Bruno and C.Chappert, Phys.Rev.Lett.
{\bf 67}, 1602, (1991); E.Bruno and B.L.Gyorffy, J.Phys.
Condens. Matter {\bf 5}, 2109, (1993).

\end{references}
\end{document}